\documentclass[
showkeys,
 amsmath,amssymb,
 aps,
Prb,
]{revtex4-2}

\usepackage{graphicx}
\usepackage{dcolumn}
\usepackage{bm}
\usepackage{color}
\usepackage{subfigure}

\begin{document}

\preprint{APS/123-QED}

\title{Enhanced Sensitivity and Wave-Structure Interaction in Nonsingular Flat-Band Lattices with Compact Localized States }

\author{Emanuele Riva}
\email[]{emanuele.riva@polimi.it}
\author{Jacopo Marconi}
\author{Francesco Braghin}
\affiliation{Department of Mechanical Engineering, Politecnico di Milano, 20156 (Italy)}

\date{\today}

\begin{abstract}
This paper investigates the dynamics of compact localized modes in one-dimensional flat-band elastic lattices. Flat dispersion arises from destructive interference between neighboring elements, resulting in a zero group velocity across all momenta. This unique condition enables the formation of wave modes that are not only highly localized in space and inherently non-propagative—protected by the flatness of the dispersion relation—but also exceptionally sensitive to structural variations due to enhanced wave-structure interaction. These features are first explored on a simple spring-mass lattice and later applied to a microelectromechanical (MEMS) system of oscillators. By exploring the role of flat-band dispersion in mechanics, this work provides new insights into their fundamental dynamics while opening new opportunities for applications in vibration control and the sensitivity analysis of mechanical structures.


\end{abstract}


\maketitle

\section{Introduction}
Wave propagation in architected materials has seen groundbreaking advancements in recent years, with phononic crystals and acoustic metamaterials driving the discovery of novel wave phenomena through carefully engineered dispersion characteristics \cite{oudich2023tailoring}. 
In this context, topological mechanics has emerged as a powerful framework for wave manipulation, where complex geometries and symmetry-breaking mechanisms are harnessed to generate localized modes that guide waves along boundaries and domain walls, including those with sharp curves. These effects draw direct analogies to the quantum Hall \cite{chen2019mechanical}, quantum spin Hall \cite{chen2018elastic}, and quantum valley Hall effects \cite{riva2018tunable,ni2017topological,ma2019topological,ganti2020topological}, which are particularly advantageous mechanisms for their back-scattering immunity and robustness against defects—attributes rooted in the nontrivial topological properties encoded in reciprocal-space. 
Another approach to topology-based design exploits virtual dimensions in parameter space, which enables the exploration of higher-dimensional topological effects within lower-dimensional systems. For instance, edge states typically associated with 2D quantum Hall effect (QHE) systems have been demonstrated in 1D periodic \cite{martinez2019edge,rosa2019edge,riva2021adiabatic} and quasiperiodic systems \cite{kraus2012topological,ni2019observation,pal2019topological,xia2020topological,pantaleoni2024topological}. Within this framework, topological pumping emerges as a mechanism that induces edge-to-edge and edge-to-interior modal transitions, driven by parametric variations along an additional dimension, either spatial \cite{riva2020edge} or temporal \cite{riva2020adiabatic,xia2021experimental,grinberg2020robust}, thereby guiding vibrations between distinct lattice positions. All these examples showcase how localized modes facilitate waveguiding and vibration control in mechanical systems.

A distinct approach to wave manipulation leverages the customization of dispersion properties through carefully designed unit cell parameters. One notable strategy involves the use of nonlocal interactions, where elastic connections extending beyond nearest neighbors are tailored to achieve specific dispersion characteristics \cite{braghini2021non,rosa2020dynamics}. This approach has enabled the realization of unique phenomena, such as maxon and roton dispersion in one-dimensional lattices \cite{PhysRevLett.131.176101}, and partially flat dispersion bands in two-dimensional systems \cite{PhysRevB.110.144304}. Alternatively, graded metamaterials leverage spatially or temporally varying elastic properties to achieve gradual, adiabatic modifications of the dispersion landscape. This capability facilitates the control and manipulation of elastic waves, including wavenumber or frequency transformations. As a result, graded metamaterials can be employed to achieve a variety of effects, such as wave trapping \cite{riva2024adiabatic,meng2020rainbow}, steering \cite{santini2023elastic,xu2019deflecting}, or mode conversion \cite{SANTINI2024118632,IORIO2024118167}.

Driven by the potential of localized modes to enable dispersion-tailored wave propagation and vibration control, this paper explores a new line of work to achieving flat dispersion through destructive interference, a phenomenon reminiscent of geometric frustration, commonly observed in magnetism, cold atomic systems, and photonic lattices \cite{rhim2021singular}. This phenomenon is herein explored through a simple spring-mass lattice in a triangular configuration and later extended to a system of microelectromechanical oscillators (MEMS) with the same triangular geometry. The specific arrangement of the lattice sites, whether on spring-mass or MEMS lattices, induces destructive interference between neighboring elements, resulting in a flat dispersion band that is well separated from the other bands. This unique condition facilitates the formation of compact localized states (CLSs)—wave modes where energy remains confined to a small region, with amplitudes dropping to zero outside this localized zone \cite{bergman2008band,rhim2021singular,maimaiti2017compact}. 
Notably, CLSs arise in both nonsingular and singular configurations. In nonsingular configurations, the Bloch bands are well-separated, preserving distinct Bloch modes. Conversely, singular flat bands exhibit band crossings where the algebraic and geometric multiplicities differ, giving rise to closed-loop modes that can be selectively formed through linear combinations of CLSs \cite{rhim2021singular,emanuele2024creating}. This paper focuses on nonsingular configurations, where CLSs are utilized to confine elastic waves in an extremely localized manner while maintaining sustained oscillations over time. Under this condition, the lattice exhibits enhanced wave-structure interaction, akin to the light-matter interactions observed in photonic flat-band systems \cite{gersen2005real,baba2008slow,krauss2007slow}. Such interactions significantly enhance sensitivity to structural changes, including linear eigenfrequency shifts in response to the variation of system parameters and eigenvector localization in spatial correlation with the parameter variation. These features are explored in the paper, providing new insights into the fundamental dynamics of flat-band lattices while unveiling the potential of these systems for highly sensitive structural diagnostics and wave control.

The structure of the article is as follows: Section II covers the theoretical foundations, including the dispersion characteristics and the existence of compact localized states (CLS) in elastic lattices. Section III presents simulations in the time and frequency domains. Section IV discusses the implementation in a MEMS structure. Finally, Section V provides concluding remarks.

\section{Theory}
\begin{figure}[!t]
    \centering
    {\includegraphics[width=0.8\textwidth]{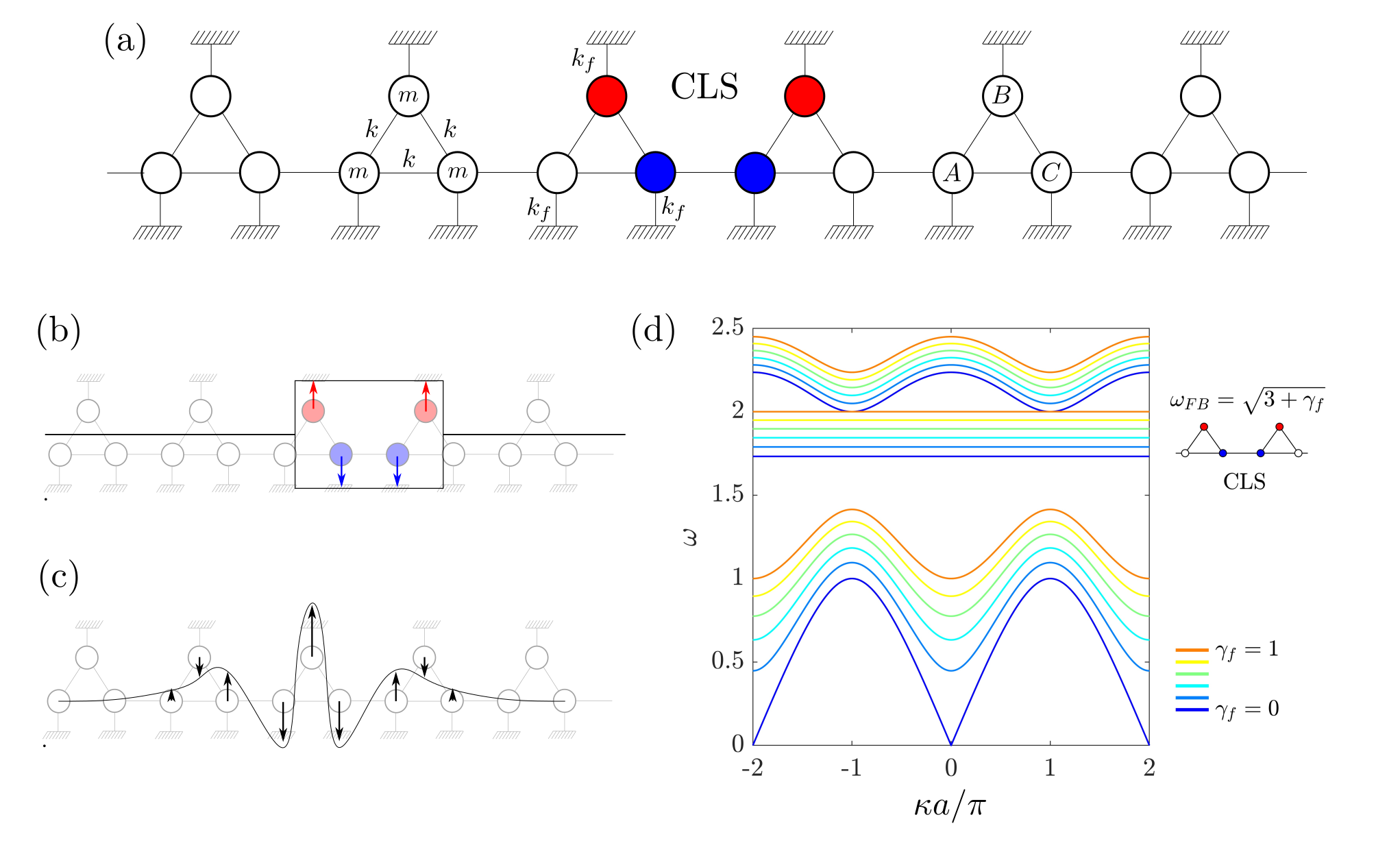}}
    \centering
    \caption{(a) Schematic of the spring-mass system. (b) Graphical representation of a compact localized state (CLS) and (c) graphical representation of a localized mode (defect mode or interface mode). (d) Dispersion relation of the lattice upon varying the ground stiffness ratio $\gamma_f$, where the schematic of the CLS and the flat dispersion equation are reported alongside the dispersion diagram.}
    \label{fig:01}
\end{figure}
The aim of this section is to characterize the dynamics of compact localized states (CLSs) for the lattice depicted in Fig. \ref{fig:01}(a). 
The system, inspired by prior works on photonic lattices \cite{ramezani2017non}, comprises $S=6$ unit cells arranged in a triangular configuration, where each mass $m$ is connected to the others via a stiffness $k$ and to the ground through a stiffness $k_f$. A finite lattice strip is formed by connecting neighboring elements A and C using the same stiffness $k$ as employed for the internal connections. 
This condition enables destructive interference when the energy is properly injected into the system, which is a phenomenon inherently connected to the existence of CLSs \cite{rhim2021singular}. CLSs are extremely localized modes in space, as illustrated in Fig. \ref{fig:01}(b). Unlike defect modes or topological modes, which are represented schematically in Fig. \ref{fig:01}(c), CLSs occupy a finite spatial region and exhibit exact zero amplitude outside this localized zone, thereby promoting a much greater degree of localization. 
The existence of these intriguing modes is demonstrated starting with the elastodynamic equation governing the $n^{th}$ unit cell:
\begin{equation}
\begin{split}
    m\ddot{u}_A^{(n)}-ku_C^{(n-1)}+\left(3k+k_f\right)u_A^{(n)}-ku_B^{(n)}-ku_C^{(n)}=0\\[5pt]
    m\ddot{u}_B^{(n)}-ku_A^{(n)}+\left(3k+k_f\right)u_B^{(n)}-ku_C^{(n)}=0\\[5pt]
    m\ddot{u}_C^{(n)}-ku_A^{(n-1)}+\left(3k+k_f\right)u_C^{(n)}-ku_B^{(n)}-ku_A^{(n+1)}=0
\end{split}
\end{equation}
where $u_C^{(n-1)}=u_C^{(n)}{\rm e}^{{\rm i}{\kappa}\cdot{a}}$ and $u_A^{(n+1)}=u_A^{(n)}{\rm e}^{-{\rm i}{\kappa}\cdot{a}}$, being $\kappa$ the wavenumber and $a$ the primitive vector. Defining $\gamma_f=k_f/k$ and $\Omega=\omega/\omega_0$, with $\omega_0=\sqrt{k/m}$, along with a harmonic motion $u_{A,B,C}=\hat{u}_{A,B,C}{\rm e}^{{\rm i}\omega t}$, the solution of the eigenvalue problem $\kappa\left(\Omega\right)$ determines the dispersion relation of the system:
\begin{equation}
    \begin{bmatrix}
        &3+\gamma_f-\Omega^2&-1&-\left(1+{\rm e}^{{\rm i}{\kappa}\cdot{a}}\right)\\[5pt]
        &-1&2+\gamma_f-\Omega^2&-1\\[5pt]
        &-\left(1+{\rm e}^{-{\rm i}{\kappa}\cdot{a}}\right)&-1&3+\gamma_f-\Omega^2
    \end{bmatrix}\begin{pmatrix}
        \hat{u}_A^{(n)}\\[5pt]
        \hat{u}_B^{(n)}\\[5pt]
        \hat{u}_C^{(n)}\\
    \end{pmatrix}=0
\end{equation}
which leads to the following expression for the three dispersion curves: 
\begin{equation}
    \Omega_1=\sqrt{\gamma_f+\frac{5}{2}-\sqrt{\frac{25}{4}+2\left(\cos\kappa a-1\right)}}\hspace{1cm}    \Omega_2=\sqrt{3+\gamma_f}\hspace{1cm}    \Omega_3=\sqrt{\gamma_f+\frac{5}{2}+\sqrt{\frac{25}{4}+2\left(\cos\kappa a-1\right)}}
\end{equation}
where the second band is independent of $\kappa$ and, hence, flat over the Brillouin Zone, as shown in the dispersion diagram in Fig.\ref{fig:01}(d) for different values of $k_f$. Without any loss of generality, $a=1$, $k=1$, $\gamma_f=0.5$, and $m=1$ are employed in the remainder of the paper. The discussion is now focused on the flat dispersion band $\Omega_2$, where the eigenvector takes the following form:
\begin{equation}
    \left|\mathbf{u}\right>=\begin{pmatrix}
        {\rm e}^{{\rm i}{\kappa}\cdot{a}}\\
        -1-{\rm e}^{{\rm i}{\kappa}\cdot{a}}\\
        1\\        
    \end{pmatrix}
\end{equation}
Note that, in contrast to singular configurations, the eigenvector does not vanish and, after normalization, is not discontinuous both at the origin and also for any other momenta within the Brillouin zone \cite{emanuele2024creating,rhim2021singular}:
\begin{equation}
        \left|\mathbf{u}_\kappa\right>=\frac{\left|\mathbf{u}\right>}{\alpha_\kappa}
\end{equation}
where $\alpha_k=\langle\mathbf{u}|\mathbf{u}\rangle=4+2\cos{\kappa a}$. Now, the construction of a CLS is accomplished by taking the linear combination of Bloch wave functions populating the flat-band:
\begin{equation}    \left|\chi_{\bm{R}}\right>=\sum_{\bm{\kappa}\in BZ}\displaystyle\alpha_{\bm \kappa}{\rm \displaystyle e}^{-{\rm i}\bm{\kappa}\cdot\bm{R}}\left| \bm{\psi}\left(\bm{\kappa}\right)\right>
\end{equation}
where the Bloch eigenstate is $\left| \bm{\psi}\left(\bm{\kappa}\right)\right>=\sum_{\bm{R}'}\sum_{j=A,B,C}{\rm \displaystyle e}^{{\rm i}\bm{\kappa}\cdot\bm{R}'}u_j\left(\bm{\kappa}\right)\left|\bm{R}',j\right>$. Here, $\bm{R}$ is the lattice vector, i.e. the position of the unit cell, while $\left|\bm{R}',j\right>$ denotes the position of the $j^{th}$ lattice site located at $\bm{r}=\bm{R}'$, and $u_j\left(\bm{\kappa}\right)$ denotes the corresponding $j^{th}$ amplitude displacement.
\begin{figure}[!t]
    \centering
    {\includegraphics[width=1.0\textwidth]{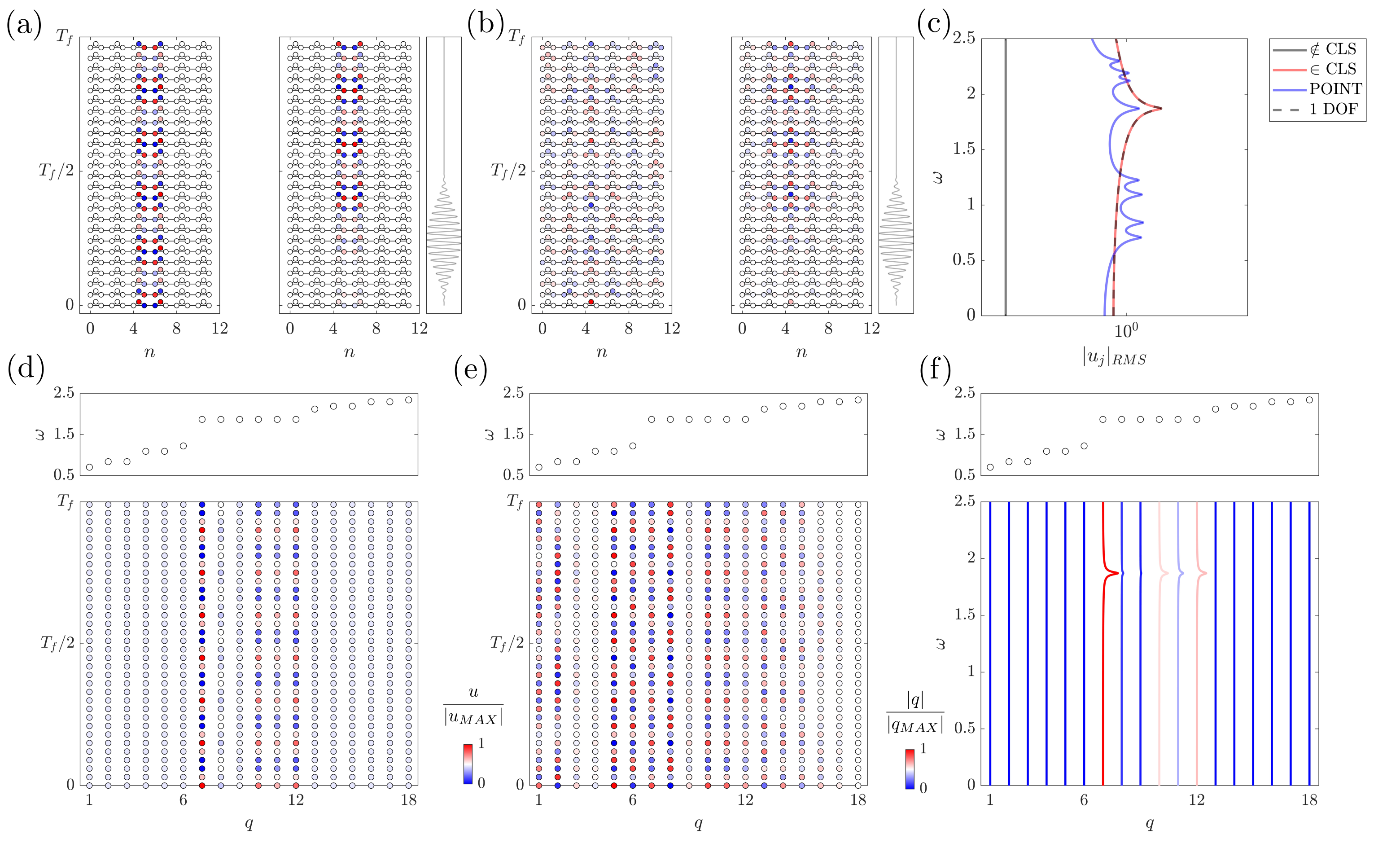}}
    \centering
    \caption{(a) Time response of the lattice for initial conditions with non-zero components corresponding to the shape of a CLS (left panel) and for a narrowband 10-period harmonic excitation centered at the flat-band frequency (right panel). (b) Time response of the lattice to point-like initial conditions applied at the third lattice element (left panel) and to a narrowband 10-period harmonic excitation applied at the same location (right panel). (c) Frequency response to CLS excitation, averaged outside the CLS domain (black line) and inside the CLS domain (red curve). The dashed line represents the response of a single degree-of-freedom (1-DOF) oscillator with a resonant frequency of $\sqrt{3+\gamma_f}$ and unitary mass. (d) Time history of the response to CLS initial conditions projected onto the generalized coordinate set $\mathbf{q}$. (e) Time history of the response to point-like initial conditions projected onto the generalized coordinate set $\mathbf{q}$. (f) Frequency response due to CLS excitation projected onto the generalized coordinate set $\mathbf{q}$. } 
    \label{fig:02}
\end{figure}
In a triangular lattice that supports flat-band characteristics, the amplitude corresponding to each element in the $j^{\text{th}}$ unit cell centered in zero, which is obtained by projecting the wave function $\left|\chi_{\bm{R}}\right>$ into $\bm{R}'$, i.e., $\left<\bm{R}',j\right|\left.\chi_{\bm{R}}\right>$, is non-zero only for a finite set of lattice vectors:
\begin{equation}
\begin{pmatrix}
\left<\bm{R}',A\right|\left.\chi_{\bm{0}}\right>\\
\left<\bm{R}',B\right|\left.\chi_{\bm{0}}\right>\\
\left<\bm{R}',C\right|\left.\chi_{\bm{0}}\right>\\
\end{pmatrix}=
\begin{pmatrix}
\delta_{R,-a}\\
-\delta_{R,0}-\delta_{R,-a}\\
\delta_{R,0}\\
\end{pmatrix}
\label{eq:07}
\end{equation}
That is, negative $-1$ for $u_B^{(n)}$ and $u_B^{(n+1)}$, positive $1$ for $u_C^{n}$ and $u_A^{n+1}$, and zero elsewhere, which corroborates the extremely localized nature of the flat-band states. The process of constructing the CLS in this way is analogous to evaluating the dynamic response at the flat-band frequency, a result that will be confirmed through numerical simulations, hereafter reported. 
\section{Results}
The presence of CLSs is now investigated using numerical simulations in both the time and frequency domains. The discussion starts by exploring the localization properties of CLSs, and later shifts focus on the wave-structure interaction. Particular attention is given to the eigenvalue sensitivity to structural changes and enhanced sensing, capabilities arising from the flat-band characteristics.  
 To study these effects, the boundaries of the system are eliminated by enforcing continuity conditions on the left and right ends of the chain. While these boundary conditions do not correspond to those of a real finite system, they enable a more symmetric distribution of states within the dispersion bands, making them more representative of the analyzed lattice. Moreover, this approach does not compromise the generality of the study, as similar outcomes can be obtained using clamped or free boundary conditions.

First and foremost, to excite a CLS it is essential to employ a forcing mechanism tailored to the form specified in Eq. \ref{eq:07}. Under such conditions, both broadband and narrowband excitations effectively generate a CLS characterized by sustained temporal oscillations, as illustrated in Fig. \ref{fig:02}(a). The left panel of the figure depicts the system response to initial conditions shaped by Eq. \ref{eq:07}, while the right panel shows the response to a 10-period harmonic excitation centered on the flat-band frequency, with the excitation applied at the same location. In both scenarios, the energy remains confined and localized without spreading to adjacent lattice sites.

In contrast, Fig. \ref{fig:02}(b) demonstrates a different outcome: point-like initial conditions and excitations result in energy transfer to neighboring elements. This process activates states with non-zero velocity, corresponding to modes from other dispersion branches. To corroborate this observation, the displacement is projected onto a set of generalized coordinates, $\mathbf{q}=\Psi^{-1}\mathbf{u}$, where $\Psi$ accommodates the eigenvectors $\psi_i$ relative to the eigenvalues $\omega_i$, solution of the eigenvalue problem $\left(K-\omega_i^2M\right)\psi_i=0$, being $M$ and $K$ the mass and stiffness matrices of the finite lattice. The results of this projection are presented in Fig. \ref{fig:02}(d) for CLS initial conditions and in Fig. \ref{fig:02}(e) for point-like initial conditions. The top box in these figures displays the distribution of natural frequencies. Notably, Fig. \ref{fig:02}(d) reveals that only flat-band modes are selectively excited in the case of a CLS input. Conversely, Fig. \ref{fig:02}(e) indicates that a point excitation triggers the activation of all modes populating the spectrum.

The response of the system is further analyzed in the frequency domain under the application of a harmonic force following the form of a CLS, i.e., according to Eq. \ref{eq:07}, and in the form of a point force. The results, shown in Fig. \ref{fig:02}(c), highlight the differences between these two excitation methods. To ease interpretation, for the case of CLS excitation, the response is averaged first outside and later inside the CLS region. The response outside the CLS, indicated by the black curve, is effectively zero, with very small values rounded to $10^{-5}$ for clarity in visualization. In contrast, the response within the CLS, shown as a red curve, exhibits a single resonance at the flat-band frequency, confirming that all energy is confined to the flat-band states. 
This is further confirmed by projecting the CLS response to the set of generalized coordinates $\mathbf{q}$, as shown in \ref{fig:02}(f), whereby only the modes populating the flat-bands are suitably combined to produce a compact localized response. 
Interestingly, the red curve matches the response of a single-degree-of-freedom (1-DOF) oscillator with a resonant frequency of $\sqrt{3+\gamma_f}$ and unitary mass, represented by the black dashed line. This equivalence underscores the localized and isolated dynamics associated with CLS excitation. Instead, a point, unitary force excitation yields a very different outcome, as illustrated by the blue curve. Here, the response is averaged over the entire structure, revealing multiple resonance peaks corresponding to the excitation of various modes. This results in a distributed system response, with energy spreading throughout the structure, thereby lacking the compact localization observed with CLS excitation.

\begin{figure}[!t]
    \centering
    {\includegraphics[width=1.0\textwidth]{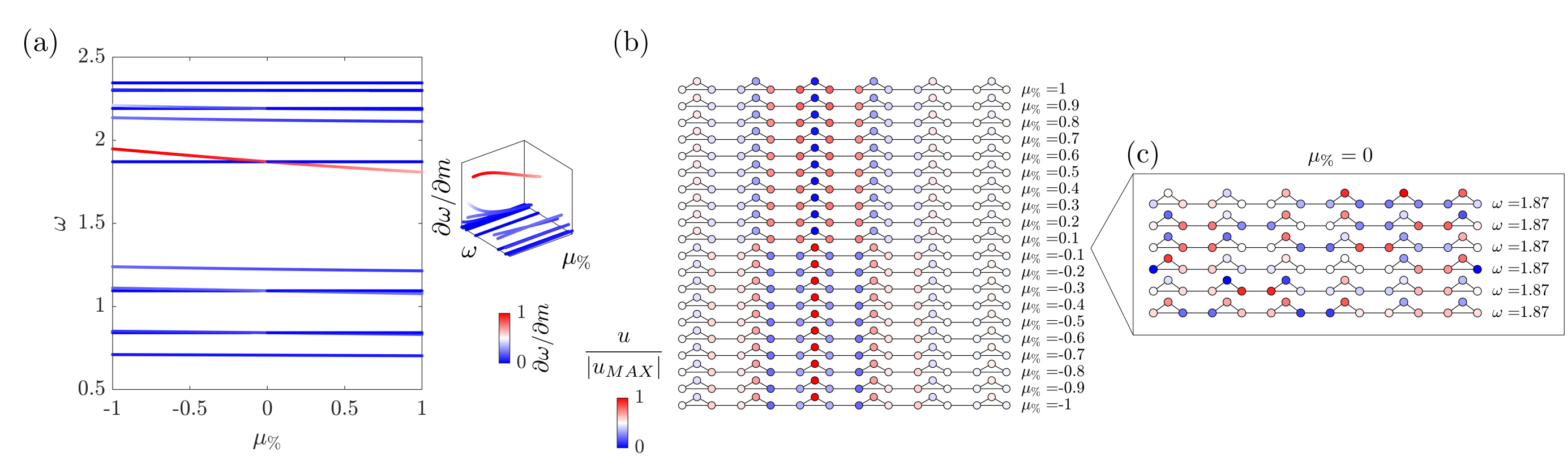}}
    \centering
    \caption{(a) Eigenvalue sensitivity as a function of the added mass ratio $\mu_\%$. (b) Eigenvector shape relative to the state separating from the flat band. (c) Graphical representation of the flat-band eigenvectors in unperturbed scenario ($\mu_\%=0$). }
    \label{fig:03}
\end{figure}

The discussion now focuses on the eigenvalue sensitivity to structural modifications. Due to the flatness of the band, the group velocity of the flat-band states vanishes, leading to enhanced wave-structure interactions. This results in significantly higher sensitivity of modes separating from the flat band to structural perturbations compared to other modes. This behavior is depicted in Fig. \ref{fig:03}(a), which presents the lattice spectrum as the mass ratio $\mu_\%=100\cdot m_{ADD}/m_{TOT}$ varies. Here, $\mu_\%$ denotes the ratio between the added mass and the total mass of the structure, which practically represents a structural defect applied to site B of the third lattice element. Note that the diagram is color-coded, where the intensity of the color reflects the level of sensitivity $\partial\omega_i/\partial m$. Furthermore, to better illustrate the variations of $\omega$ across the desired parameter space $\mu_\%$, the sensitivity $\partial\omega_i/\partial m$ is computed numerically by evaluating $\omega_i$ at discrete $\mu_\%$ values. Interestingly, a single mode separates from the flat band, demonstrating a high and nearly constant sensitivity. Instead, bulk modes occupying the first and third bands show much lower, though non-zero, sensitivity. In contrast, the five remaining flat-band states exhibit zero sensitivity, indicating that the structural modification exclusively affects the localized state separating in response to the structural change.
Indeed, Fig. \ref{fig:03}(b) shows the mode shape of the mass-sensitive mode, which becomes localized at site B of the third lattice element for various values of $\mu_\%$. This observation further confirms the strong link between the structural modification and the properties of the flat-band modes, where the band flatness promotes a highly sensitive measure of both the magnitude and the location of the structural change. 
Finally, to better qualify the relation between the CLS excitation and the spectral properties of the flat-band, the unperturbed flat-band modes are reported alongside in Fig. \ref{fig:03}(c). Interestingly, none of these modes are compact localized. However, a proper excitation mechanism, consistent with Eq. \ref{eq:07}, allows triggering a destructive superposition of modes outside the CLS domain, and a constructive superposition inside this domain, thereby offering an extraordinary degree of localization.

\begin{figure}[!t]
    \centering
    {\includegraphics[width=1.0\textwidth]{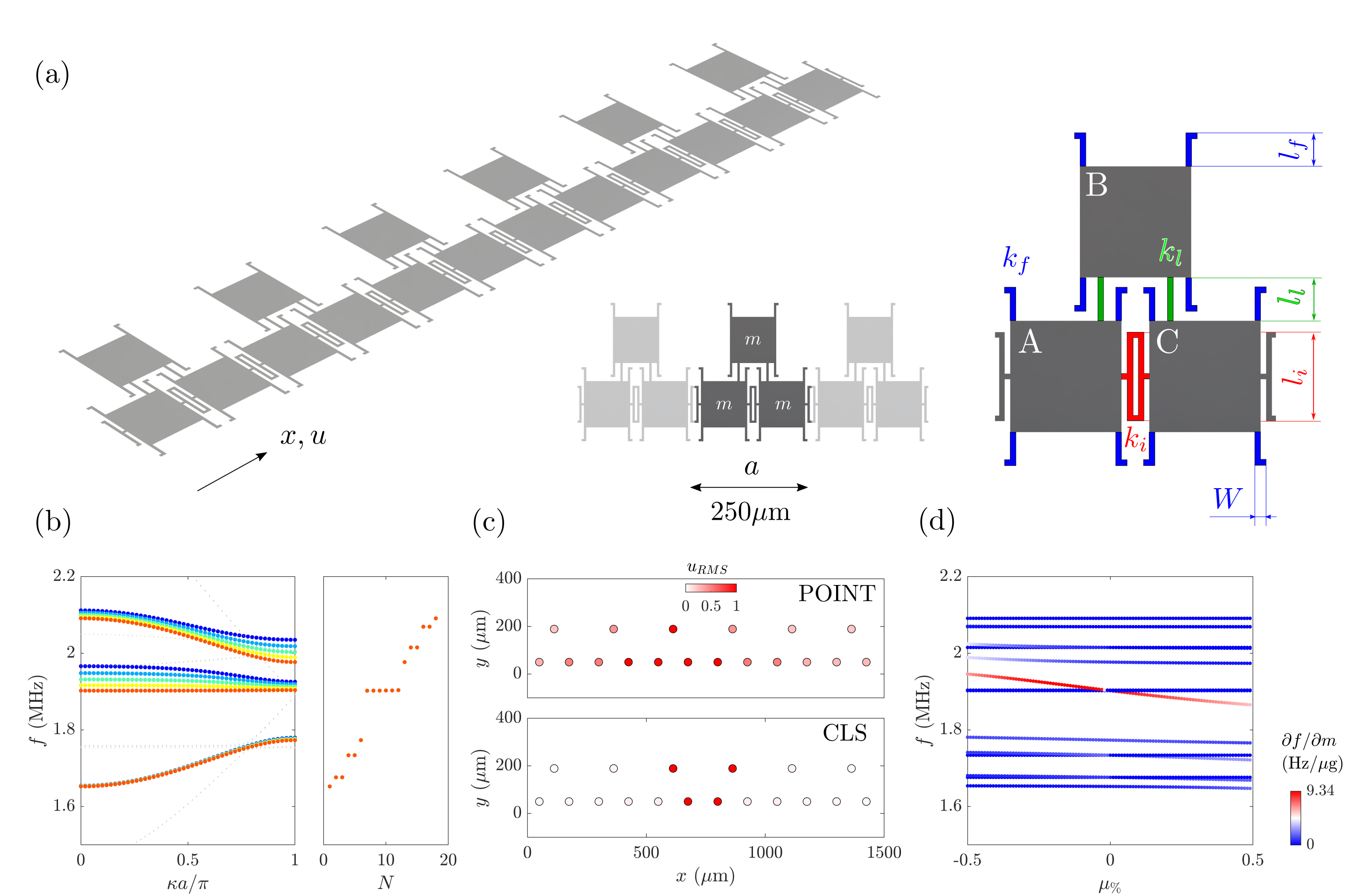}}
    \centering
    \caption{(a) schematic of the system of MEMS oscillators and zoomed view of the unit cell. (b) dispersion relation upon varying the lateral stiffness $k_l$. The eigenfrequencies of the finite lattice are reported alongside the dispersion diagram. (c) root mean square (RMS) of the time response under point excitation (top panel) and CLS excitation (bottom panel). (d) Sensitivity of the eigenvalues with respect to a change of the mass ration $\mu_\%$  }
    \label{fig:04}
\end{figure}

\section{Compact localized states and sensitivity in a system of MEMS oscillators}
The concept is now applied to a system of coupled MEMS oscillators \cite{riva2024adiabatic,marconi2023non}, which consists of $N=6$ unit cells, as illustrated in Fig. \ref{fig:04}(a). To limit the analysis to longitudinal motion, each mass is connected to the ground via blue-highlighted beams, which terminate in lateral squares that act as anchor points. Such beams serve as ground springs, rigid along the transversal direction and compliant along the longitudinal axis of the lattice. Notably, the triangular arrangement of the masses ensures the formation of compact localized states (CLSs), along with the associated wave propagation and spectral properties, provided that all the remaining elastic connections are appropriately designed. 
While the ground stiffness $k_f$ influences the flat-band frequency and the slope of the dispersion curve, it does not hinder the formation of the flat band, provided that the ground stiffnesses are uniform across all masses. For simplicity, all connections are implemented using beams with a constant width of $W=5$ $\mu$m. As such, the length of the ground spring is arbitrarily set to $l_f=30$ $\mu$m, and the internal and lateral stiffnesses $k_i$ and $k_l$ are set identical to each other to guarantee a perfect destructive interference effect. To this end, the length of the internal folded spring is set equal to $l_i=80$ $\mu$m, and a parametric sweep is applied to reach a flat dispersion, as shown in Fig. \ref{fig:04}(b), for a design length of $l_l=39$ $\mu$m. Upon application of continuity conditions to the left and right ends of the lattice, the spectrum displays $N=6$ degenerate natural frequencies in correspondence to the flat band. 

Similarly to the spring-mass lattice example, the dynamic response of the system is analyzed under point and CLS excitations, with a 10-period sinusoidal force applied at the flat-band frequency. For a point excitation, the root mean square (RMS) displacement of the lattice, shown in Fig. \ref{fig:04}(c), reveals that the energy spreads across the entire lattice. In contrast, under CLS excitation, the energy remains compactly localized across four lattice sites, as expected. Animations are also provided in the supplementary material (SM). To evaluate the sensitivity of the lattice, the mass of element B in the third unit cell is varied. The results, presented in Fig. \ref{fig:04}(d), exhibit behavior in good agreement with the spring-mass lattice example. Specifically, the mode separating from the flat band demonstrates significantly greater sensitivity compared to the global modes of the structure.

\section{Conclusions}
This study explores the dynamics of compact localized states (CLS) in flat-band lattices through numerical simulations. The existence of CLS is first analyzed theoretically and validated through both time and frequency domain analyses, demonstrating the system remarkable ability to confine elastic waves in an extremely localized manner, both in spring-mass lattices and in a system of MEMS oscillators. Additionally, the results reveal that the unique localization properties and flat-band characteristics facilitate distinctive wave-structure interactions, resulting in pronounced eigenvalue sensitivity to structural perturbations. This sensitivity establishes elastic lattices as an effective method for detecting and characterizing structural modifications via flat-band engineering.
In other words, the proposed framework provides new perspectives on wave manipulation and defect detection in flat-band systems, with promising potential applications in vibration isolation, signal filtering, and precision sensing.

\nocite{*}

\section*{References}
\bibliography{apssamp}
 
\end{document}